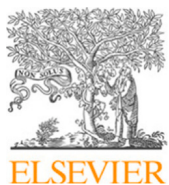
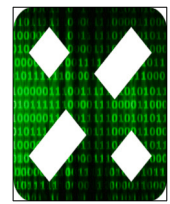

Original software publication

# Document Towers: A MATLAB software implementing a three-dimensional architectural paradigm for the visual exploration of digital documents and libraries

Vlad Atanasiu [*], Rolf Ingold

*University of Fribourg, Department of Informatics, Bd. de Pérolles 90, 1700 Fribourg, Switzerland*

**ABSTRACT**

This article introduces the generic Document Towers paradigm, visualization, and software for visualizing the structure of paginated documents, based on the metaphor of DOCUMENTS-AS-ARCHITECTURE. The Document Towers visualizations resemble three-dimensional building models and represent the physical boundaries of logical (e.g., titles, images), semantic (e.g., topics, named entities), graphical (e.g., typefaces, colors), and other types of information with spatial extent as a stack of rooms and floors. The software takes as input user-supplied JSON-formatted coordinates and labels of document entities, or extracts them itself from ALTO and InDesign IDML files. The Document Towers paradigm and visualization enable information systems to support information behaviors other than goal-oriented searches. Visualization encourages exploration by generating panoramic overviews and fostering serendipitous insights, while the use of metaphors assists with comprehension of the representations through the application of a familiar cognitive model. Document Towers visualizations also provide access to types of information other than textual content, specifically by means of their physical structure, which corresponds to the material, logical, semantic, and contextual aspects of documents. Visualization renders documents transparent, making the invisible visible and facilitating analysis at a glance and without the need for physical manipulation. Keyword searches and other language-based interactions with documents must be clearly expressed and will return only answers to questions asked; by contrast, visual observation is well suited to fuzzy goals and uncovering unexpected aspects of the data.

© 2021 The Author(s). Published by Elsevier B.V. This is an open access article under the CC BY license (http://creativecommons.org/licenses/by/4.0/).

## Code metadata

| | |
|---|---|
| Current code version | v2020.02.15 |
| Permanent link to code/repository used for this code version | https://github.com/ElsevierSoftwareX/SOFTX-D-21-00009 |
| Code Ocean compute capsule | – |
| Legal Code License | BSD-3-Clause |
| Code versioning system used | None |
| Software code languages, tools, and services used | MATLAB (R2020b) |
| Compilation requirements, operating environments & dependencies | Linux, Mac, Unix, Windows |
| If available Link to developer documentation/manual | Documentation included in software. |
| Support email for questions | atanasiu@alum.mit.edu |

## 1. Motivation, principles and significance

*"Mathematics can only become truly interesting and original when it involves the operation of seeing something as something else."*

[Reviel Netz]

* Corresponding author.
  *E-mail addresses:* atanasiu@alum.mit.edu (Vlad Atanasiu),
rolf.ingold@unifr.ch (Rolf Ingold).
  *URLs:* http://waqwaq.info/ (Vlad Atanasiu),
https://www3.unifr.ch/inf/de/all/people/16738/0a54b (Rolf Ingold).

Documents composed of pages are ubiquitous information storage formats, whether in digital (e.g., PDF and Word documents) or analog form (e.g., books, articles). Massive ongoing efforts are being made to realize the analog-to-digital conversion of historical document collections (e.g., Google Books), while digital-to-digital format conversions are also routine (e.g., Word to PDF). Thus, the relevance of developing diverse solutions that facilitate interactions with this type of information cannot be overstated. The Document Towers paradigm, visualization, and software introduced herein are one of these solutions.

While, typically, document content is accessed through the semantic analysis of text and graphical objects, this information is not always available (e.g., optical character recognition [OCR] is needed to make text in document images searchable) or may be expensive (e.g., high-quality handwriting recognition remains costly today). One solution, implemented in the Document Towers software, is to exploit the information contained in the document structure. For example, the first page of an article usually features a greater number and diversity of visually distinct entities than the subsequent pages; it also contains a greater variety of information levels, such as title, abstract, and authors. In other words, the structure of information is potentially indicative of its informativeness. It could in fact be argued that information *is* structure: letters and pixels create different texts and images depending on how they are arranged.

Natural language and numerical descriptions are other common paradigms for accessing information. The visual modality adopted by the Document Towers human–machine interface offers an alternative with several advantageous characteristics, including its support for exploration, serendipitous discoveries, and overview. These kinds of information behaviors differ from the targeted search implicit in the concept of the text input fields used almost ubiquitously in Internet search engines [1–3]. Visual search is thus well suited to cases in which the information sought cannot be clearly expressed in words, in which users employ different terminology and languages to describe the same concept, or in which the findings may be unexpected (as in quality control).

The Document Towers visualization represents documents as wireframe architectural models. The purpose of this cognitive model is to create a sense of familiarity for the user while comprehending and manipulating opaque physical objects and abstract digital structures [4,5]. There is a natural material correspondence between stacks of pages and buildings, and between libraries and cities; moreover, spatial and architectural metaphors are ingrained in software and hardware terminology (e.g., "information superhighway", "homepage", "desktop", "tunnel", "cloud", and the scientific field of "Information Architecture") [6], and have also been used in the past for experimental document representations. Examples include visualizing software structure as an urban landscape [7], extruding the nested structure of webpage objects [8], representing text columns in a three-dimensional space [9], rendering digital documents as look-alikes of physical books in a virtual reality library [10], and conceptualizing navigation between Internet domains as a walk through tunnels that connect various rooms [11]. Research into compact document overview are addressed in the literature, e.g., in the form of semantically highlighted thumbnails [12], or a dashboard of topic distribution in documents [13]. Illustrated surveys of similar experiments can be found in [14–16].

Fig. 1 illustrates how a three-dimensional architectural model is obtained by extruding the bounding boxes of various objects on a document page, thereby creating slabs that resemble rooms, walls, or pillars. Fig. 3 depicts how the page boundaries look like floors and the entire document like a building. Fig. 6 shows the similarity between a collection of documents and a cityscape.

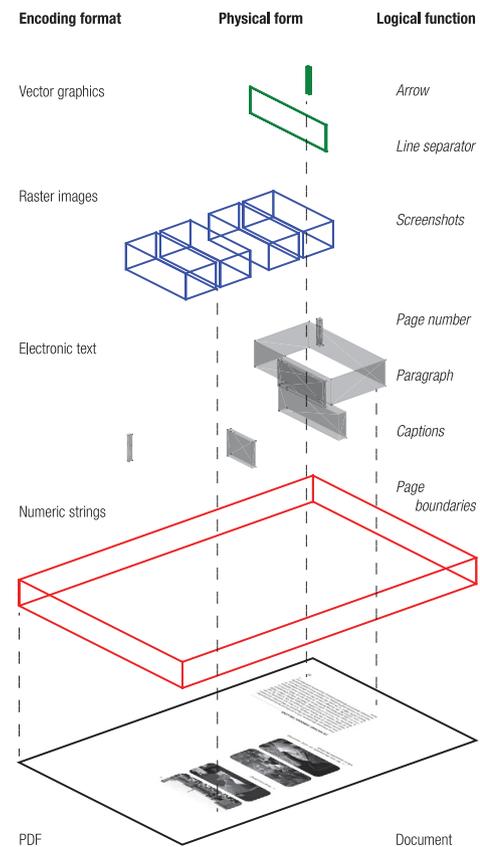

**Fig. 1.** Principle by which page entities are transformed into an architectural model.

Two effects of representing documents as wireframe architectural models contribute to the paradigm's effectiveness in eliciting insights. First, the models preserve the physical structure of paginated documents, which facilitates their analysis. Second, the simplification of shapes to slabs helps on one hand focus on the essential, and on the other hand spurs the imagination to search for meanings to sibylline structures, which is the goal of a visualization meant for exploration.

The Document Towers are first and foremost a visualization *paradigm* for paginated documents, while the Document Towers *software* implements this paradigm in an interactive visual *interface* (Fig. 2, left). Its purpose is to enable the exploration of document structures for practical applications, to aid research into the interpretation of document structures, and to promote the Document Towers paradigm for potential adoption in third-party applications. A distinct functionality of the software is its ability to read document object geometry and metadata from selected document formats, although the software scope in this case is not automatic analysis conducted by machines, but visual analysis performed by humans.

The separation between generic visualization and processing specific document formats is achieved by using as input to the software a custom JSON-formatted intermediary file for simple geometry and metadata description, which can be either generated from document files by the software or supplied by the users. To produce Document Towers visualizations all is needed are coordinates of the spatial extent of entities within documents, while the labels adding metadata to the entities are useful, but not indispensable. This makes the proposed method a generic information visualization concept for paginated documents, just like a histogram is a generic concept for representing a set of scalars.

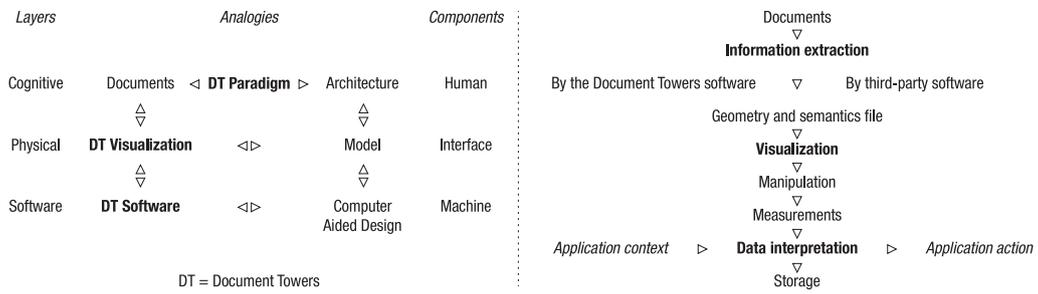

**Fig. 2.** Document Towers representation process (left) and workflow (right).

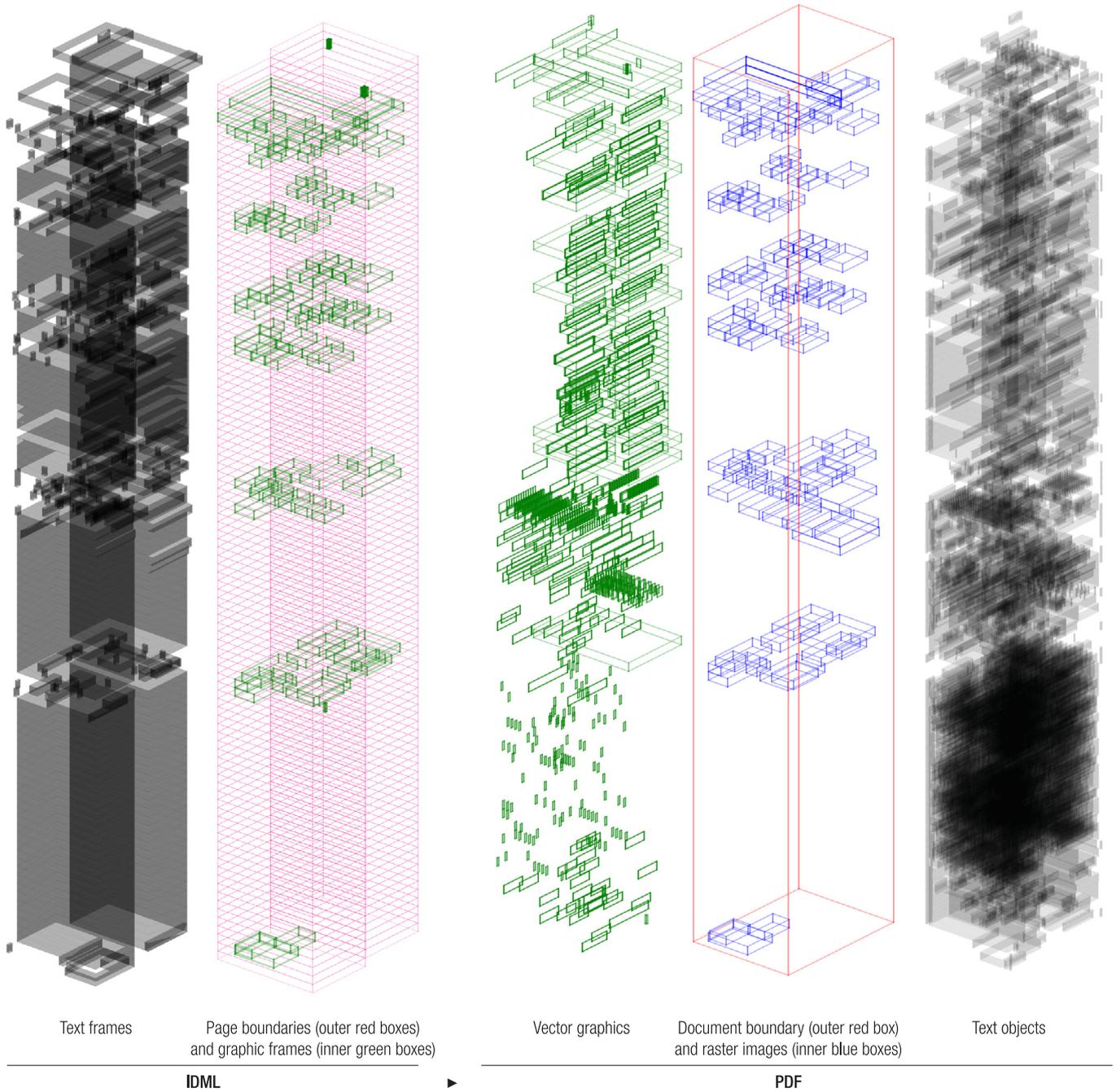

**Fig. 3.** Document Towers visualizations representing the distribution of text and graphics in the same document, generated from an IDML file (left) and a PDF file (right). (For a color version of this figure, the reader is referred to the web version of this article.)

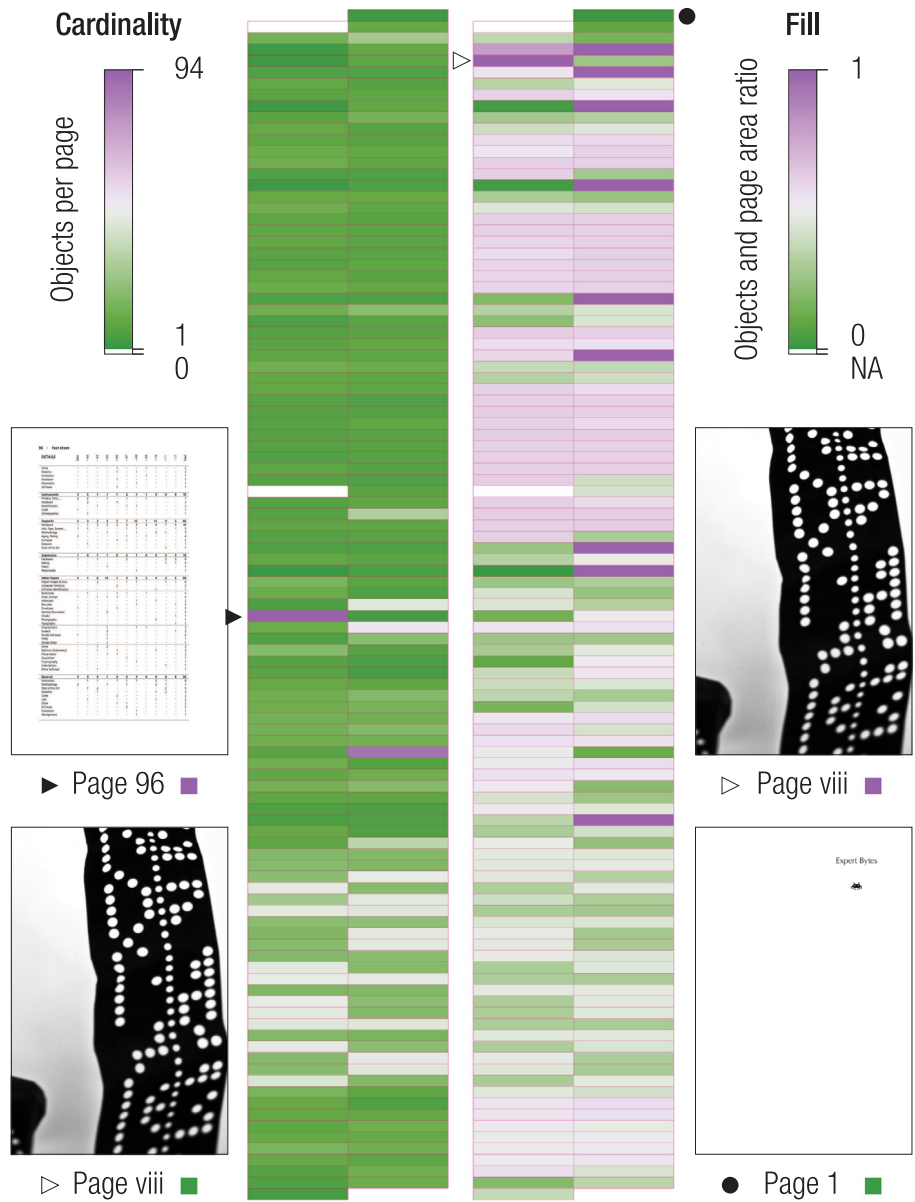

**Fig. 4.** Measures and Ribbons. Please refer to Section 2.6 "Measurements" for details. (For a color version of this figure, the reader is referred to the web version of this article.)

## 2. Software description

### 2.1. Software architecture and workflow

The software consists of a data capture and processing module, an interactive document structure viewer, and feature measurement functions. The user begins by using the software interface to select one or more electronic document files containing coordinates of document objects, and possibly also labels specifying the object classes (Fig. 2, right). The software extracts the object geometry and labels, then saves them to a file. This file can also be produced by a third-party software package in cases where it is necessary to process formats not supported by the Document Towers software (see the software documentation for specifications). In the next step, the user selects these geometry files, after which the software displays three-dimensional wireframes at the specified locations, and colors them according to the labels. The user can now interactively explore and interpret the visualization, which can subsequently be saved for later reuse.

### 2.2. Programming environment

The Document Towers software is written in the MATLAB programming language (R2020b). The software can be used in conjunction with MathWorks' MATLAB commercial application, or distributed for free as a compiled standalone desktop application running on top of the royalty-free MATLAB Runtime Compiler [17].

The visual quality of the 3D wireframes was first tested in Java (JavaFX), but early releases proved inadequate [18]. A JavaScript three.js library for 3D animation and using WebGL was found to be both graphically satisfactory and appealing due to its allowing for Web delivery of the visualization [19]. However, the exceedingly poor performance when importing geometry files larger than 1 MB made this solution impractical. For reference, the uncompressed ASCII file size for the geometry of a single Tower in Fig. 3 is 343 kB (4265 objects), while the collection of 89 Document Towers visualizations in Fig. 6 is 6.7 MB (85 203 objects).

Creating Document Towers visualizations in MATLAB resulted in none of the graphical drawbacks of JavaFX and was reasonably fast: a single Tower can be generated in four seconds on a MacBook Pro 2018 (3 GHz Intel Core i9). Use of the game engine Unity [20] and the design software AutoCAD [21] was contemplated, but MATLAB has the advantage of being a scientific software development environment that supports a broad range of applications, including image processing and statistics, both of which are useful for extending the document analysis power of the Document Towers software.

### 2.3. Extracted information

The information extracted from document files comprises the coordinates of the spatial extent of entities within the document, along with labels defining these entities (when available). For example, the IDML and ALTO files discussed in the next section provide labels for four spatially defined basic document entities: pages, text frames, raster images, and vector graphics. ALTO files typically contain a further set of labels, created ad hoc by the file producers, which introduce subcategories for the basic document entities (such as "title" for a text frame, "stamp" for a raster image, and color values for a vector font). The graphical interface of the Document Towers software can be used to filter entities for display. The Document Towers visualizations are agnostic to what they represent; their power depends on the richness of the information in various document formats, the possibility of automatically extracting it, and the user's ability to interpret it [22].

### 2.4. Data formats

The software is capable of reading the object coordinates and labels present in XML-formatted IDML (InDesign Markup Language) [23] and ALTO (Analyzed Layout and Text Object) [24] files. The former is employed by Adobe InDesign, the market-dominant typography software used by the publishing industry for complex layouts, while the latter is a standard for digitization projects.

Coordinates within pages, page numbers, object types, and labels, along with document metadata such as file names, are saved to a file in JSON (JavaScript Object Notation) format [25]. As an example, the first line in the file describing the Document Tower in Fig. 1 is [0, 0, 0, 441, 0, 441, 666, 0, 666]: here, the initial zero defines the object as a page, while the following numbers are the coordinate pairs of its bounding box. Sample data and detailed formatting specifications are provided with the software.

The geometry file is subsequently read by the visualization module. This hand-over mechanism allows the user to create document structure libraries, split long documents, merge documents into a single Document Tower, modify labels and metadata, or process data structure files generated using third-party software. For example, the information extracted from PDFs (Portable Document Format) presented in Figs. 3 and 6 is obtained through a proprietary application programming interface (API) for the Enlighter software of the Swiss company Sugarcube [26].

There may well be substantial differences in the information obtained from the various document formats; as a result, the visualized structures can be quite distinct from each other. Taking as an example an academic book written and produced in InDesign by the first author [27], Fig. 3 illustrates how the IDML-based visualization (left) emphasizes a recurrent irregular layout in the first part of the document, resulting from the presence of illustrations, while the Tower derived from the PDF of the same document (right) highlights the stark paragraph density of the reference section at the document's end. Many of the vector graphics visible in the PDF are rulers, which separate text from footnotes or table sections; in the IDML, these are defined as paragraph *parameters*, not as identifiable *objects*.

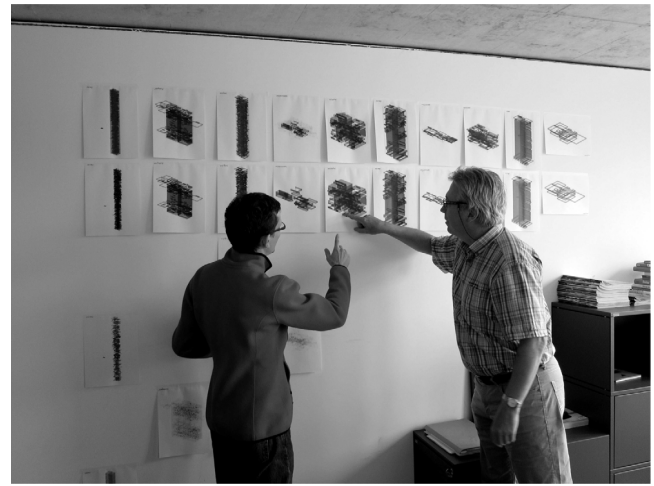

**Fig. 5.** The authors in front of a Document Towers visualization mural.

These differences raise questions, such as that of how much uncertainty related to document layout is present in a particular document description file. For example, a LATEX .tex file on its own reveals little about the final appearance of the document, while a PDF file is expected to provide faithful graphical representation. Rather than considering the data format-dependency phenomenon a limitation, it may be useful to instead interpret differences as providing valuable insights into the specificities of various document formats, some of which may become visible through the Document Tower visualization paradigm. It is also useful to consider the complementarity of document formats, and investigate, via the Document Towers visualizations as thinking tools, how the different information provided by these formats might be used individually or fused. Such questions are relevant to document engineering, archival, and forensic applications, and the Document Towers paradigm provide a perspective distinct from others to address these questions.

### 2.5. Visualization and interaction

The main document structure representation paradigm implemented by the Document Towers visualization is that of three-dimensional wireframes that utilize the architectural metaphors of OBJECTS-AS-ROOMS, PAGES-AS-FLOORS, DOCUMENTS-AS-TOWERS, and LIBRARIES-AS-CITIES. By color-coding the facade of each floor according to specific semantic and quantitative criteria, so-called Ribbons are obtained, which are a more space-saving representation than the Document Towers visualization (Fig. 4).

Through a graphical interface, parameters such as colors and transparency can be set, the type of objects to display can be selected, and the projection changed between axonometric, elevation, and plan view; the user can also zoom, pan, and rotate the Document Towers visualizations in order to move from overview to details. Hyperlinks can be added, enabling a PDF version of the document to be opened at desired pages in a web browser; in this way, the Document Towers visualizations acquire a document navigation function similar to a table of contents.

### 2.6. Measurements

The software provides numerical statistics on object categories, as well as colorcoded information (Ribbons) on the number of objects per page (*cardinality*), and on the percentage of the area of a page covered by objects (*fill*). Fig. 4 presents a front view of a Document Tower from the same PDF file as in Fig. 3, in which the facade was colorcoded on two measurements with reference to text, vector, and raster objects. The pages corresponding to the extreme values are shown. Page 96 is a table and has the highest number of objects per page as extracted by the PDF analysis algorithm, while page viii is a single raster image that covers the entirety of the page surface, in contrast to the almost empty page 1.

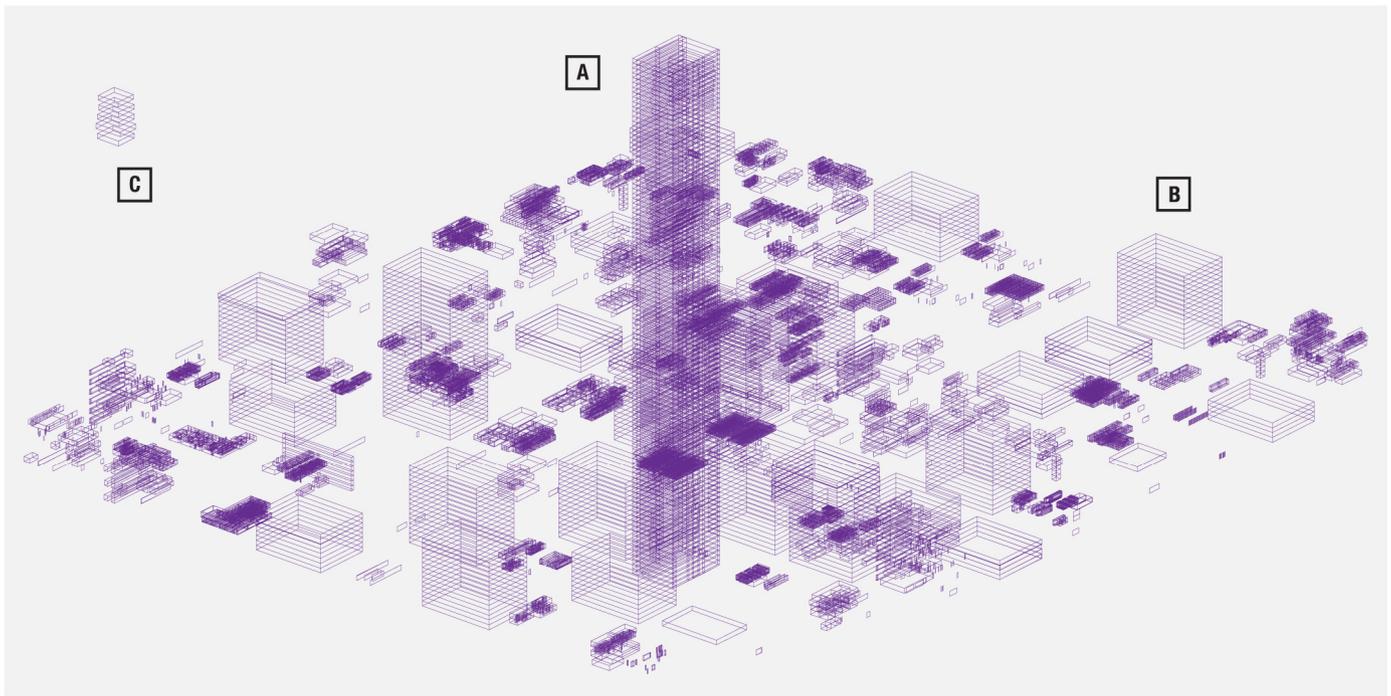

**Fig. 6.** A Document Towers City visualization having lead to three insights discussed in Section 3 "Illustrative example".

*2.7. Scalability*

The scalability of rendering documents as Document Towers visualizations is constrained by performance and physical factors. The former was discussed in Section 2.2 "Programming environment", while the latter derives from the limited surface and resolution available on typical computer displays, restricting the detail and overview capacity of representations. About 3000 article-length documents such as in Fig. 6 might be shown with reasonable legibility on the 3072 × 1920-pixels screen of a MacBook Pro laptop. To represent more documents at higher resolutions, it is suggested to make hard copies and display them on walls, a technique used by graphic designers for magazines, and architects and engineers for plans. This approach was also adopted by the authors to compare document structures, yielding additional benefits of persistent visualizations and more space to accommodate a larger group of viewers (Fig. 5).

**3. Illustrative example**

The Document Towers visualizations shown in Fig. 6 represent eighty-nine PDF documents in a folder ordered by filename on one of the authors' computers; here, the blue slabs represent the location and extent of raster images in the digital documents. The image exemplifies the following three benefits of visualizing document structures. All of these are serendipitous discoveries made by visualizing the document with the Document Tower software.

*A. Misclassification:* While this collection was supposed to contain only articles, the presence of a high-rising Tower reveals a monograph among them. This misclassification would most likely have gone unnoticed if not for its serendipitous discovery due to visualization.

*B. Search quality:* The regular Document Towers visualizations represent scanned documents, where each page is a single raster image, while the fragmented Document Towers visualizations represent native electronic documents, in which images cover only a portion of pages (if present at all). For a library wishing to offer its readers searchable digital documents, the implication is that scanned documents must be identified and the text and logical structure extracted, which is time-consuming and costly. Moreover, the text recognition rate from raster images rarely yields an electronic text identical to the one in the imaged documents, particularly in the case of noisy, historical, or handwritten documents, which leads to suboptimal textual search results. Visualization, by contrast, offers a quick and lightweight means to estimate document search quality before performing text recognition.

*C. Forensics:* The small Tower in the upper left corner, which draws the eye due to its outlying location, is not an individual document, but rather represents images located outside the visible frame of the PDF documents. Could it conceal a hidden message?

**4. Impact**

The Document Towers paradigm combines document structure with an architectural metaphor. This paradigm has a potentially broad range of both applications and users, and covers the entire document lifecycle.

Its main impact pertains to the information-seeking strategy it fosters: namely, exploration as opposed to targeted search. This is of particular interest in rich, unstructured, and undocumented environments, such as archives.

Document digitization and conversion, in particular, are applications likely to benefit from visual document structure representation, as this method is cheap to implement (e.g., can be applied before optical character recognition) and provides rich insights through non-linguistic means by leveraging human pattern recognition capabilities (helpful for, e.g., outlier detection in quality control).

The Document Towers visualizations also make information in documents visible without the need to open them. As such, they could act as document navigation aids (e.g., next to the table of contents in e-book readers). They might further help designers to overview documents they are laying out.

Last but not least, Document Towers visualizations have a certain aesthetic appeal, which is a useful user experience factor [28].

## 5. Conclusions

The article introduced the Document Towers paradigm, visualization, and software, designed to facilitate the visual exploration of paginated document structures using an architectural cognitive model. Beyond the specifics of the visualization and software, the usefulness of the work lies in the paradigm it promotes: that of a technological information-seeking aid complementary to linguistic and numerical targeted search.

## CRediT authorship contribution statement

**Vlad Atanasiu:** Conceptualization, Methodology, Software, Writing – original draft, Visualization. **Rolf Ingold:** Conceptualization, Writing – review & editing, Supervision, Funding acquisition.

## Declaration of competing interest

The authors declare that they have no known competing financial interests or personal relationships that could have appeared to influence the work reported in this paper.

## Acknowledgments

The first author expresses gratitude for the support and stimulating discussions with Andreas Fischer of the University of Fribourg, and Jean-Luc Bloeche and Maurizio Rigamonti of Sugarcube. Both authors acknowledge the improvement brought by the gracious comments of the anonymous reviewers. K. F. is thanked for proofreading. This research was supported by the Fund in Support of Innovation of the canton of Fribourg, Switzerland, under grant No. 2013.03.